\documentclass[useAMS,usenatbib]{mn2e}
\usepackage{epsfig}
\usepackage{color} 
\usepackage{natbib}
\usepackage{deluxetable}
\usepackage{amssymb}
\usepackage{rotating}
\usepackage{graphicx}
\usepackage[colorlinks=true,citecolor=blue]{hyperref}
\usepackage{mathpazo}
\usepackage{graphicx}
\usepackage{footnote}
\usepackage{longtable}
 \usepackage[american]{babel}
\title[Optical monitoring of 3  $\gamma$-ray NLSy1s]
{Intra-Night Optical monitoring of three $\gamma$-ray detected Narrow-line Seyfert 1 galaxies}
\author[$Ojha$, $Gopal-Krishna$ \& $Chand$ ]{Vineet Ojha$^{1}$\thanks{E-mail: vineet@aries.res.in
}, Gopal-Krishna$^{1}$, Hum Chand$^{1}$ \\ $^{1}$Aryabhatta Research Institute of Observational Sciences (ARIES),
  Manora Peak, Nainital $-$ 263002, India\\}

\begin{document}
\date{Accepted ---. Received ---; in original form ---}

\pagerange{\pageref{firstpage}--\pageref{lastpage}} \pubyear{2017}

\maketitle

\label{firstpage}
\begin{abstract}
For 3 radio-loud $\gamma$-ray detected Narrow-Line Seyfert 1 ($\gamma$-ray NLSy1) galaxies,
we report optical variability on intra-night and/or week-like time scales, based
on five $\geq$ 3 hours long monitoring sessions for each galaxy. The
radio-loudness factors (R$_{1.4 GHz}$)$\textcolor{red}{^{1}}$ for these galaxies, namely 1H 0323$+$342 (z
= 0.0629), PKS J1222$+$0413 (z = 0.966) and PKS J1505$+$0326 (z = 0.408)
 are $\sim$318, $\sim$1534 and $\sim$3364 at 1.4 GHz, respectively. For
the most distant $\gamma$-ray NLSy1, PKS J1222$+$0413, Intra-Night Optical Variability (INOV) characterisation
is presented for the first time. The blazar-like behaviour of the nearest $\gamma$-ray NLSy1
1H 0323$+$342, which showed strong INOV on 4 of the 5 nights, was
unexpected in view of its recent reclassification as ‘radio intermediate’ (R$_{5 GHz}$
$\lesssim$ 25). Its particularly violent INOV is manifested by 
two optical outbursts lasting $\sim$ 1 hour, whose rapid
brightening phase is shown to imply a doubling time of $\sim$ 1 hour for
the optical synchrotron flux, after (conservatively) deducting the
thermal optical emission contributed by the host galaxy and the
Seyfert nucleus. A more realistic ‘decontamination’ could well reduce
substantially the flux doubling time, bringing it still closer in rapidity
to the ultra-fast VHE ($>$ 100 GeV) flares reported for the blazars PKS
1222$+$216 and PKS 2155$-$304. A large contamination by thermal optical emission
may, in fact, be common for NLSy1s as they are high Eddington rate
accretors. The present study further suggests that superluminal motion in the radio jet could be a robust
diagnostic of INOV.

\end{abstract}



\section{Introduction}
\label{Sect 1}
Intensity variation over the entire accessible electromagnetic
spectrum is one of the defining characteristics of active galactic
nuclei (AGN). This trait is often utilised as an effective tool to
probe their emission mechanism on physical scales that are
inaccessible to direct imaging
techniques~\citep[e.g.,][]{Wagner-Witzel1995ARA&A..33..163W,
  Urry1995PASP..107..803U, Ulrich1997ARA&A..35..445U,
  Zensus1997ARA&A..35..607Z}. The optical flux variations of AGN
occurring on hour-like, or occasionally even shorter timescales are
commonly known as Intra-Night Optical Variability (INOV) and it has
come to be used quite extensively as a tracer of jet activity in
blazars and other AGN
classes~\citep[e.g.,][]{Miller1989Natur.337..627M,Gopal-Krishna1993A&A...271...89G,Gopal-Krishna1995MNRAS.274..701G,
  Jang1995ApJ...452..582J,Heidt1996A&A...305...42H,
  Bai1999A&AS..136..455B,Romero1999A&AS..135..477R,
  Fan2001A&A...369..758F, Stalin2004MNRAS.350..175S,
  Gupta2005A&A...440..855G,
  Carini2007AJ....133..303C,2009AJ....138..991R,
  Goyal2012A&A...544A..37G, Goyal2013MNRAS.435.1300G,
  Kumar2017MNRAS.471..606K}.\par In the case of blazars, the INOV
phenomenon is usually associated with Doppler boosting of the
jet's radiation, which not only amplifies any emission fluctuations occurring
within the jet's plasma whose bulk relativistic motion is directed close
to our line of sight but also shortens the timescales~\citep[e.g.,][]{Hughes1992ApJ...396..469H,
  Marscher1992vob..conf...85M, Begelman2008MNRAS.384L..19B,
  Ghisellini2008MNRAS.386L..28G, Giannios2009MNRAS.395L..29G,
  Marscher2014ApJ...780...87M}. At a subdued level, the same process
may be at work in radio-quiet quasars
(RQQs) due to the presence of a weak jet~\citep{Gopal-Krishna2003ApJ...586L..25G,
  Stalin2004MNRAS.350..175S, Barvainis2005ApJ...618..108B}, although hot spots or flares on
accretion discs may also be significant, if not the dominant
contributor to their INOV~\citep{Mangalam1993ApJ...406..420M, Wiita2006ASPC..350..183W}. For radio-quiet\footnote{Radio-loudness
  is usually parametrised by the ratio (R) of the rest-frame flux
  densities at 5 GHz and at 4400\AA, being R$\leq$ and $>$ 10 for
  radio-quiet, radio-loud quasars,
  respectively~\citep[e.g.,][]{Kellermann1989AJ.....98.1195K}. To differentiate the radio-loudness estimates based on the flux
densities at 1.4 GHz and 5 GHz, we have used R$_{1.4 GHz}$ and R$_{5 GHz}$ notations, respectively.}  AGN
showing rapid X-ray variability, such as Narrow-Line Seyfert 1
galaxies (NLSy1s), one may also expect to find short-term optical
variations simply because the X-ray emission may have an optical tail.
However, any such evidence is
weak~\citep[e.g.,][]{Miller1996ASPC..110...17M,
  Ferrara2001ASPC..224..319F}, although at longer time scales this
phenomenon is well known~\citep[e.g.,][]{Rokaki1993A&A...272....8R,
  Gaskell2006ASPC..360..111G}. \par

Fairly extensive information is now available on the INOV properties
of luminous AGN for both radio-quiet and radio-loud varieties,
including blazars, as summarized recently
by~\citet{Gopal-Krishna2018BSRSL..87..281G}. Much more scarce,
however, are INOV data for their low-luminosity counterparts, e.g.,
NLSy1s~\citep[e.g.,][and references
  therein]{Kshama2017MNRAS.466.2679K}.  NLSy1s are characterised by
the narrow width of optical Balmer emission lines FWHM(H${\beta}) <
2000$  km s$^{-1}$~\citep{Osterbrock1985ApJ...297..166O,
  Goodrich1989ApJ...342..908G, Pogge2000NewAR..44..381P,
  Sulentic2000ApJ...536L...5S}, a small flux ratio 
[O$_{III}]_{\lambda5007}/H\beta<$
3~\citep{Shuder-Osterbrock1981ApJ...250...55S}.  With some possible
exceptions, they also exhibit strong [Fe VII] and [Fe X]
lines~\citep{Pogge2011nlsg.confE...2P}, as well as strong permitted
optical/UV Fe~{\sc ii} emission
lines~\citep{Boroson1992ApJS...80..109B, Grupe1999A&A...350..805G}. As
a class, NLSy1s are hosted by spiral
galaxies~\citep[e.g.,][]{Crenshaw2003AJ....126.1690C,
  Deo2006AJ....132..321D}, although early-type galaxies have been
considered as the host for the radio/$\gamma$-ray loud subset of
NLSy1s~\citep{2008A&A...490..583A,
  2014ApJ...795...58L}. Interestingly, their soft X-ray emission which
has a steep spectrum~\citep{Boller1996A&A...305...53B,
  Wang1996A&A...309...81W, Grupe1998A&A...330...25G}, is often rapidly
variable~\citep{Leighly1999ApJS..125..297L,
  Komossa-Meerschweinchen2000A&A...354..411K,
  Miller2000NewAR..44..539M}.  At optical wavelengths, the first
reports of rapid variability of NLSy1s on hour-like time  scales
appeared almost two decades ago~\citep{Miller2000NewAR..44..539M,
  Ferrara2001ASPC..224..319F, Klimek2004ApJ...609...69K}.  A major
boost to the studies of NLSy1 galaxies is likely to come from the
recent publication of a large sample of 11,101 NLSy1s, out of which
$\sim$ 600 are radio loud (\citealp[see
  also]{Rakshit2017ApJS..229...39R},
\citealp[]{Chen2018arXiv180107234C};
\citealp[]{Singh2018arXiv180701945S}).\par From the analysis of
optical spectroscopic data, it has been inferred that the central
black holes in NLSy1s have virial masses mostly in the range 10$^{6}-
10^{8} M_{\sun}$~\citep[e.g.,][]{Mathur2000MNRAS.314L..17M,
  Peterson2000ApJ...542..161P, Yuan2008ApJ...685..801Y,
  Xu2012AJ....143...83X, Foschini2015A&A...575A..13F,
  Cracco2016MNRAS.462.1256C}. Thus, they are typically 1-2 orders of
magnitude less massive than the black-holes embedded in the cores of
broad-line Seyfert galaxies and more powerful radio sources, like
blazars and radio-loud quasars, which are nearly always hosted by
early-type galaxies~\citep[e.g.,][]{McHardy1994MNRAS.268..681M,
  Boyce1998MNRAS.298..121B, Scarpa2000ApJ...544..258S,
  Boroson2002ApJ...565...78B,2016MNRAS.460.3202O}, and have black-hole
masses above 10$^{8}
M_{\sun}$~\citep[e.g.,][]{Laor2000ApJ...543L.111L,
  Dunlop2003MNRAS.340.1095D, McLure2004MNRAS.353L..45M,
  Chiaberge2005ApJ...625..716C, Gopal-Krishna2008ApJ...680L..13G,
  Chiaberge2011MNRAS.416..917C, Tadhunter2016A&ARv..24...10T,
  Coziol2017MNRAS.466..921C}.  For  $\gamma$-ray NLSy1s,
somewhat higher BH masses (between a few 10$^{7}$ and a few 10$^{8}
M_{\sun}$) have been derived by modeling the optical/UV part of the
SED in terms of a Shakura \& Sunyaev
disc~\citep[e.g.,][]{Doi2012ApJ...760...41D,
  Calderone2013MNRAS.431..210C, Foschini2015A&A...575A..13F,
  D'Ammando2016MNRAS.463.4469D, Paliya2016ApJ...820...52P} and they
lie at the lower end of the BH masses of quasars/blazars. Physical
scenarios for the possibility that virial masses of NLSy1s black holes
may have been underestimated, include the effect of radiation
pressure~\citep{Marconi2008ApJ...678..693M}, and a pole-on view of a
disk-like BLR~(\citealp[e.g.,][]{Nagao2001ApJ...546..744N};
\citealp{Bian2004MNRAS.352..823B};
\citealp{Decarli2008MNRAS.386L..15D};
\citealp{Shen2014Natur.513..210S}; \citealp{Baldi2016MNRAS.458L..69B};
\citealp[see, however,][]{Jarvela2017A&A...606A...9J}).\par

The existence of the above mass discrepancy came into the spotlight
following the discovery of $\gamma$-ray emission from a few NLSy1s,
using the {\it
  Fermi}/LAT\footnote{https://heasarc.gsfc.nasa.gov/docs/heasarc/missions/fermi.html}~\citep{Abdo2009ApJ...699..976A,
  Abdo2009ApJ...707..727A, Abdo2010ApJ...716..835A,
  Calderone2011MNRAS.413.2365C, Foschini2011MNRAS.413.1671F,
  D'Ammando2012MNRAS.426..317D, D'Ammando2015MNRAS.452..520D,
  Yao2015MNRAS.454L..16Y, Paliya2018ApJ...853L...2P}. Nearly 20 NLSy1
galaxies have since been catalogued as $\gamma$-ray emitters and all
of them are also radio detected~\citep[e.g.,][and references
  therein]{Berton2018arXiv180508534B, Paliya2018ApJ...853L...2P}.
Their detection at radio and $\gamma$-ray bands has reinforced the
view that in spite of being hosted by spiral galaxies, their central
engines are capable of ejecting relativistic jets emitting strong
nonthermal radiation, a hallmark characteristic of
blazars~\citep[e.g.,][]{Yuan2008ApJ...685..801Y,
  Foschini2015A&A...575A..13F}. Their similarity to blazars is
bolstered due to detection of the double-humped SED profile for
several $\gamma$-rays NLSy1s ~\citep[e.g.,][]{Abdo2009ApJ...699..976A,
  Abdo2009ApJ...707..727A, Foschini2011MNRAS.413.1671F}, in both
flaring and non-flaring
states~\citep[e.g.,][]{D'Ammando2016MNRAS.463.4469D,
  Paliya2016ApJ...820...52P}. We note, however, that very recent
studies have revealed that weak relativistic radio jets may even be
launched by non-blazar type NLSy1
galaxies~\citep{2018A&A...614L...1L}.\par

Purely from the radio perspective, there is a clear evidence for
(quasar-like) bi-modality in the radio loudness of NLSy1s. However,
the radio-loud fraction is smaller; $\leq$ 7\%  NLSy1s have a
radio-loudness parameter  R$_{5 GHz} >$
10~\citep[e.g.,][]{Komossa2006AJ....132..531K,
  Zhou2006ApJS..166..128Z, Rakshit2017ApJS..229...39R,
  Singh2018arXiv180701945S}. This fraction is very similar to the
radio-loud fraction  of 4.7\% (taking R$_{5 GHz} >$ 10) estimated
by~\citet{Rafter2009AJ....137...42R} for a flux-limited subset of  5477
broad-line AGN drawn from a low-z sample of AGNs, extracted
by~\citet{Greene2007ApJ...667..131G} from the SDSS/DR4~\citep{
  York2000AJ....120.1579Y, Adelman-McCarthy2006ApJS..162...38A}.
In several radio-loud NLSy1s,
kiloparsec-scale radio emission has been
detected~\citep[e.g.,][]{Doi2012ApJ...760...41D,
  Foschini2015A&A...575A..13F, Congiu2017A&A...603A..32C,
  Singh2018arXiv180701945S}, although their flat spectrum subset
exhibits very dim diffuse radio
emission~\citep{Congiu2017A&A...603A..32C,
  Berton2018arXiv180508534B}.\par

Even prior to the {\it Fermi}/LAT discovery of (variable) $\gamma$-ray
emission, the flat/inverted radio spectra, high radio brightness
temperatures, superluminally moving radio knots in the
VLBI images of several NLSy1s, had become powerful indicators of
relativistic jets in their
cores~\citep[e.g.,][]{Zhou2003ApJ...584..147Z, Doi2006PASJ...58..829D,
  Yuan2008ApJ...685..801Y, D'Ammando2013MNRAS.433..952D}.  Raising the
radio-loudness threshold to  R$_{5 GHz}>$ 100, which is probably a more secure
criterion for radio
loudness~\citep[e.g.,][]{Falcke1996ApJ...471..106F,
  Rafter2011AJ....141...85R}, the radio-loud fraction of NLSy1s drops
to just 2-3\%~\citep{Komossa2006AJ....132..531K}.  It is interesting
to recall that~\citet{Zhou2007ApJ...658L..13Z} have argued that most
of such `very radio-loud' NLSy1s are in fact `radio-intermediate' AGN
with Doppler boosted nuclear radio emission. For quasars, the
intermediate range of  R$_{5 GHz}$ (10 -- 100) has long been associated with
`radio intermediate' classification and they have even been postulated
to be the tiny subset of normal QSOs whose intrinsically weak
relativistic jets appear Doppler boosted as they happen to be pointed
close to our direction~\citep{Miller1993MNRAS.263..425M,
  Falcke1996ApJ...473L..13F, Wang2006ApJ...645..856W}. Irrespective of
the underlying physical mechanism, it is a subset of clearly
radio-loud and  $\gamma$-ray NLSy1 galaxies, which will be the
focus of the present study. Specifically, we shall present the results
of our intranight optical monitoring (15 nights) of 3 such NLSy1
galaxies, having large R$_{1.4 GHz}$ (Table~\ref{tab:source_info}), all of which
are also confirmed $\gamma$-ray
emitters~\citep{Abdo2009ApJ...707L.142A,Yao2015MNRAS.454L..16Y}. In
recent years it has often been pointed out for NLSy1s that their
optical, near-infrared and even radio flux variability is similar to
blazars~\citep[e.g.,][]{Liu2010ApJ...715L.113L,
  Jiang2012ApJ...759L..31J, Paliya2013MNRAS.428.2450P,
  Angelakis2015A&A...575A..55A}. The 3 NLSy1s discussed in this paper
constitute the radio loudest subset of the 25 NLSy1s which we are
currently monitoring for intra-night and longer-term optical variability. They were extracted from
the~\citet{Foschini2011nlsg.confE..24F} sample of 76 NLSy1s confirmed
to be emitters of high-energy radiation: X-rays (detected with
ROSAT\footnote{{https://heasarc.gsfc.nasa.gov/docs/rosat/rosat.html}})
and/or $\gamma$-rays (detected with {\it Fermi}-LAT). This paper is
structured as follows. In Sect.~\ref{Sect 2} we describe our observations and
the data reduction procedure, while Sect.~\ref{Sect 3} presents the statistical
analysis of the light curves. Our main results and discussion are
presented in Sect.~\ref{Sect 4}, followed by conclusions in Sect.~\ref{Sect 5}.

\begin{table*}
 \begin{minipage}{500mm}
 
\caption{Basic parameters of the three $\gamma$-ray NLSy1 galaxies and of their central engines.
\label{tab:source_info}}
  {\small
\begin{tabular}{lcccccccccc}
  \hline

\multicolumn{1}{c}{Name (SDSS name)}  &  {R.A.(J2000)} & {Dec.(J2000)}  & $z$   &  m$_{B}$    & $\alpha_{rad}^{a}$ &   R$_{1.4 GHz}^{b}$ & M$_{BH}^{c}$ & $\lambda_{Edd}$$^{d}$  &$^{e}\Gamma_{SED}$ & PM(yr$^{-1}$)$^{f}$\\     
\multicolumn{1}{c}{}       &  (h m s)    &($^\circ$ $^\prime$ $^{\prime\prime}$) &  (redshift)    &   (mag)   &    & & (10$^{7} M_{\sun}$)&    &     & (15 GHz) \\
\multicolumn{1}{c}{(1)}   &   (2)          &                (3)                    &   (4)           &    (5)    & (6)&  (7)                  & (8) & (9) &(10) &(11)       \\

\hline
 1H 0323$+$342 (J032441.20$+$341045.0) & 03 24 41.20  &$+$34 10 45.00 & 0.063 & 16.38&$+$0.1   & 318    & (1-3)      &0.4 &7-8$^{*}$   &9.0c$\pm$ 0.3c  \\
 PKS J1222$+$0413 (J122222.99$+$041315.9) & 12 22 22.99  &$+$04 13 15.95 & 0.966 & 17.88&$+$0.3   & 1534   & 20 &0.6  &30 &0.9c$\pm$ 0.3c \\
 PKS J1505+0326 (J150506.48$+$032630.8) & 15 05 06.48  &$+$03 26 30.84 & 0.408 & 18.99&$+$0.3   & 3364   & 4      &0.1 &17    &1.1c$\pm$ 0.4c   \\

\hline

\multicolumn{11}{l}{$^{a}$Radio spectral index (S$_{\nu} \propto \nu^{\alpha}$) values for  1H 0323$+$342, PKS J1222$+$0413 and PKS J1505+0326 are taken from } \\
\multicolumn{11}{l}{~\citet{Neumann1994A&AS..106..303N},~\citet{White1992ApJS...79..331W} and ~\citet{Angelakis2015A&A...575A..55A}, respectively.}\\
\multicolumn{11}{l}{$^{b}R_{1.4 GHz}$ ($f_{1.4GHz}/f_{4400\AA}$) values for  1H 0323$+$342 and PKS J1505$+$0326 are taken from ~\citet{Foschini2011nlsg.confE..24F} and}\\
\multicolumn{11}{l}{ for PKS J1222$+$0413, R$_{1.4 GHz}$ value is estimated taking its core radio flux density of 0.6 Jy at 1.4 GHz}\\
\multicolumn{11}{l}{~\citep{Kharb2010ApJ...710..764K} and  f$_{\nu}$(4400\AA) from~\citet{Yao2015MNRAS.454L..16Y}.}\\
\multicolumn{11}{l}{ $^{c}$Black hole masses for  1H 0323$+$342,  PKS J1222$+$0413 and are PKS J1505$+$0326 taken from~\citet{Zhou2007ApJ...658L..13Z},}\\
\multicolumn{11}{l}{~\citet{Yao2015MNRAS.454L..16Y} and ~\citet{Paliya2016ApJ...820...52P}, respectively.}\\
\multicolumn{11}{l}{ $^{d}$Eddington ratios for  1H 0323$+$342,  PKS J1222$+$0413 and  PKS J1505$+$0326 are taken from ~\citet{Paliya2014ApJ...789..143P},}\\
\multicolumn{11}{l}{~\citet{Yao2015MNRAS.454L..16Y} and ~\citet{D'Ammando2016MNRAS.463.4469D}, respectively.}\\
\multicolumn{11}{l}{$^{e}$The bulk Lorentz factors $(\Gamma_{SED})$ for  1H 0323$+$342,  PKS J1222$+$0413 and  PKS J1505$+$0326 are taken from~\citet{Paliya2014ApJ...789..143P}, }\\
\multicolumn{11}{l}{ ~\citet{Yao2015MNRAS.454L..16Y} and ~\citet{D'Ammando2016MNRAS.463.4469D}, respectively. The range (marked by $^{*}$) encompasses the average and active}\\
\multicolumn{11}{l}{ states of $\gamma$-ray emission, with the higher value for the active state ~\citep{Paliya2014ApJ...789..143P}.}\\
\multicolumn{11}{l}{ $^{f}$The VLBI radio knots’ proper motion measurements are from~\citet{Lister2016AJ....152...12L}. For 1H 0323+342,~\citet{Fuhrmann2016RAA....16..176F},}\\
\multicolumn{11}{l}{reported its VLBI radio knots’ proper motions to be up to $\sim$ 7c.}\\

\end{tabular}
   }  
 \end{minipage}
\end{table*}

\section{Observations and Data Reduction}
\label{Sect 2}

\subsection{Photometric Intranight Observations}
\label{Sect 2.1}
The 3 NLSy1 targets were monitored in the Johnson-Cousin R (hereafter
R$_{c}$) filter using the 1.3m telescope (DFOT) of the Aryabhatta
Research Institute of Observation Sciences (ARIES), India, located at
Devasthal, Nainital~\citep{Sagar2010ASInC...1..203S}. DFOT is a fast
beam (f/4) optical telescope with Ritchey-Chretien (RC) optics. It has
a pointing accuracy better than 10 arcsec rms. The telescope is
equipped with a 2k$\times$2k deep thermo-electrically cooled (to about
$-85^{\circ}$C) Andor CCD camera with a pixel size of 13.5 microns and
a plate scale of 0.53 arcsec per pixel, covering a FOV of $\sim$ 18
arcmin$^{2}$ on the sky. It has a readout speed of 1 MHz and a system
rms noise and gain of 7.5 $e^-$ and 2.0 $e^-$ ADU$^{-1}$,
respectively. We monitored each NLSy1 galaxy on 5 nights for $\geq$
3.0 hours, except for a slight shortfall in the duration, occurring in
the case of the NLSy1 PKS J1222$+$0413 on 28.01.2017 due to poor
weather conditions (Table~\ref{NLSy1:tab_result}). In order to get a
reasonable SNR for each photometric measurement, the exposure times
were typically set between 4 and 15 minutes, depending on the
brightness of the source and the transparency and  brightness of the sky.

\subsection{Data Reduction}
\label{Sect 2.2}
The pre-processing of the raw images (bias subtraction, flat-fielding
and cosmic-ray removal) was done using the standard tasks in the Image
Reduction and Analysis Facility {\textsc IRAF} \footnote{{Image
    Reduction and Analysis Facility (http://iraf.noao.edu/)}}. The
instrumental magnitudes of the NLSy1 and stars in the image frames
were determined by aperture photometry~\citep{1987PASP...99..191S,
  1992ASPC...25..297S}, using the Dominion Astronomical Observatory
Photometry \textrm{II} (DAOPHOT II algorithm)\footnote{{Dominion
    Astrophysical Observatory Photometry
    (http://www.astro.wisc.edu/sirtf/daophot2.pdf)}}. A crucial
parameter for the photometry is the radius of the aperture which
determines the S/N ratio of the photometric data points for a given
target. As suggested by~\citet{Howell1988AJ.....95..247H}, the S/N
ratio is maximised when the aperture radius approximately equals the
full width at half maximum (FWHM) of the point-spread function (PSF)
for the image (and decreases for both larger and smaller
apertures). In order to find an optimum aperture, we have carried out
aperture photometry, taking four aperture radii $=$ FWHM, $2 \times$
FWHM, $3 \times$ FWHM and $4 \times$ FWHM. For each CCD frame, the
value of FWHM (i.e., seeing disk radius) was determined by taking the
mean over $5$ fairly bright stars registered in the CCD
frame. Although the photometric estimates using the different aperture
radii were generally found to be in good agreement (see, also
Sect.~\ref{Sect 4.1}), the highest S/N was almost always found when the aperture
radius was set equal to $2\times$ FWHM, which was hence adopted for
the final analysis.  As pointed out
by~\citet{Cellone2000AJ....119.1534C}, contamination from the host
galaxy of the target AGN may result in spurious variability as the
seeing disk varies, specially when the aperture is small. In the
present sample, this issue is relevant in the case of the NLSy1
1H 0323$+$342 (z = 0.0629) and has been specifically addressed in
Sect.~\ref{Sect 4.2}.  In our analysis, we first found for a given session, the
median of the FWHMs measured for all the CCD frames acquired in that
session and used two times this median value as the aperture radius
for the entire session. To derive the Differential Light Curves (DLCs)
of a given target NLSy1, we selected two steady comparison stars
(designated S1 and S2) present within all the CCD frames, such that
they are close to the target NLSy1, both in location and apparent
magnitude. We were able to ensure that at least one comparison star is
within $\sim$ 1 magnitude of the target NLSy1. The parameters of
comparison stars selected for each session are given in
Table~\ref{tab_cdq_comp}. Note that the $g-r$ color difference for the
target `NLSy1' and the corresponding comparison stars is always $<$
0.80 and $<$ 1.80 with the median values of 0.42 and 1.20,
respectively (column 7, Table~\ref{tab_cdq_comp}). Analysis
by~\citet{Carini1992AJ....104...15C}
and~\citet{Stalin2004MNRAS.350..175S}, has demonstrated that color
difference of this magnitude should produce a negligible effect on the
DLCs, as the atmospheric attenuation changes during a monitoring
session. \\

\begin{table*}
\begin{minipage}{500mm}
  {
  {\small
    \caption {Basic observational parameters of the comparison stars (S1, S2) used for the three $\gamma$-ray NLSy1 galaxies.}
     
\label{tab_cdq_comp}

\begin{tabular}{lcccccr}\\

\hline
                        & & & & &   &   \\
{Name} &Dates of observations  &   {R.A.(J2000)} & {Dec.(J2000)} & { g} & { r} & { g-r}  \\
           &                    &   (h m s)       &($^\circ$ $^\prime$ $^{\prime\prime}$)  & (mag)   & (mag)   & (mag) \\
{(1)}     &  {(2)}        & {(3)}           & {(4)}                              & {(5)}   & {(6)}   & {(7)}   \\
\hline
\multicolumn{7}{l}{}\\

1H 0323$+$342&    22,23 Nov.; 02 Dec. 2016; 03, 04 Jan. 2017       &03 24 41.20  &$+$34 10 45.00 & 14.50 & 13.70& $^{*}$0.80\\   

S1  &                                                                      &03 24 53.68  &$+$34 12 45.62 & 15.60 & 14.40& $^{*}$1.20 \\
S2  &                                                                      &03 24 53.55  &$+$34 11 16.58 & 16.20 & 14.40& $^{*}$1.80 \\

PKS J1222$+$0413&      03, 04, 28 Jan.; 21, 22 Feb. 2017             &12 22 22.99  &$+$04 13 15.95 & 17.02 & 16.80& 0.22\\   

S1  &                                                                      &12 22 34.02  &$+$04 13 21.57 & 18.63 & 17.19& 1.44 \\ 
S2  &                                                                      &12 21 56.12  &$+$04 15 15.19 & 17.22 & 16.78& 0.44 \\

PKS J1505$+$0326&    25 March, 12, 21 April 2017; 11, 20  May 2018    & 15 05 06.48  &$+$03 26 30.84 & 18.64 & 18.22& 0.42 \\

S2      &                                                                  & 15 05 32.05  &$+$03 28 36.13 & 18.13 & 17.64& 0.49\\
S3      &                                                                  & 15 05 14.52  &$+$03 24 56.17 & 17.51 & 17.14& 0.37\\

\hline

\multicolumn{7}{l}{$^{*}$Due to unavailability of SDSS (g-r) colors, (B-R) colors are taken from USNO-A2.0 catalog~\citep{Monet1998AAS...19312003M}.}\\

\end{tabular}
}
  }
  
\end{minipage}
\end{table*}

\begin{table*}
 \centering
 \begin{minipage}{500mm}
 {\small
   \caption{Observational details and the INOV results for the three $\gamma$-ray NLSy1s monitored in 15 sessions.}
 \label{NLSy1:tab_result}
 \begin{tabular}{ccc ccccc cc}\\
   \hline
   {NLSy1} &{Date} &  {T}  & {N} & { F$^{\eta}$ values$^{a}$} & {INOV status$^{b}$}  &{$\sqrt { \langle \sigma^2_{i,err} \rangle}$} & $\overline{\psi}_{s1, s2}$\\
   (SDSS Name)     & dd.mm.yyyy & hr &Points in DLC  & {$F_1^{\eta}$}, {$F_2^{\eta}$} & F$^{\eta}$-test   & (q-s) & (\%)\\
   {(1)}&{(2)} & {(3)} & {(4)} & {(5)} & {(6)} & {(7)} & {(8)}&&  \\
   \hline
   
   1H 0323$+$342  & 22.11.2016 & 4.42 & 56& 1.68 ,  2.80 &  PV, V   & 0.007 & 4.5\\  
                       & 23.11.2016 & 4.27 & 54& 3.18 ,  4.31 &  V,  V   & 0.006 & 4.7\\
                       & 02.12.2016 & 4.41 & 44& 11.00, 11.50 &  V,  V   & 0.008 & 9.34\\
                       & 03.01.2017 & 3.00 & 39& 1.03 ,  1.16 &  NV, NV  & 0.009 & --\\
                       & 04.01.2017 & 3.39 & 33& 3.50 ,  3.47 &  V,  V   & 0.009 & 6.8\\

 PKS J1222$+$0413 & 03.01.2017 & 3.52 & 17&0.53, 0.25  &  NV, NV  & 0.020 & --\\  
                       & 04.01.2017 & 3.62 & 17&0.31, 0.36  &  NV, NV  & 0.014 & --\\
                       & 28.01.2017 & 2.45 & 23&0.56, 0.52  &  NV, NV  & 0.022 & --\\
                       & 21.02.2017 & 4.44 & 41&0.74, 0.76  &  NV, NV  & 0.020 & --\\ 
                       & 22.02.2017 & 5.50 & 50&3.68, 3.50  &  V , V   & 0.018 & 13.2\\ 
                       
 PKS J1505$+$0326 & 25.03.2017 & 5.34 & 42&0.62, 0.63  &  NV, NV  & 0.029 & --\\
                       & 12.04.2018 & 3.97 & 22&0.38, 0.51  &  NV, NV  & 0.040 & --\\ 
                       & 21.04.2018 & 5.18 & 25&0.50, 0.50  &  NV, NV  & 0.038 & --\\
                       & 11.05.2018 & 3.16 & 14&1.80, 1.45  &  NV, NV  & 0.037 & --\\
                       & 20.05.2018 & 3.30 & 15&0.60, 0.37  &  NV, NV  & 0.045 & --\\
                                                    
   \hline
   
   \multicolumn{10}{l}{$^{a}$The entries in the columns 5, 6, 7 and 8 correspond to an aperture radius of 6$\times$FWHM in the case of J0324$+$3410 (Sect.~\ref{Sect 4.1})}\\
   \multicolumn{10}{l}{and 2$\times$FWHM for the remaining two  NLSy1s.}\\
   \multicolumn{10}{l}{$^{b}$V=variable, i.e., confidence level $>$ 0.99; PV=probable variable, i.e., 0.95 - 0.99 confidence level; NV=non-variable, }\\
\multicolumn{10}{l}{i.e., confidence level $<$ 0.95.}\\
 \end{tabular}  
 }              
 \end{minipage} 
\end{table*}

\section{STATISTICAL ANALYSIS OF THE DLCs}
\label{Sect 3}
Since the comparison stars are close in magnitude to the target NLSy1
in practically all the cases, as mentioned above (see,
Table~\ref{tab_cdq_comp}), the effects due to the difference in the
photon noise are quite small.  We shall, therefore, use the
$F^{\eta}-$test ~\citep{Diego2010AJ....139.1269D} to check for the
presence of INOV in the DLCs, as discussed
in~\citet{Goyal2012A&A...544A..37G}.  A specific advantage of this
choice is that the results of our analysis for the NLSy1 galaxies can
be readily compared with the INOV characterisation for other major
classes of AGN, which has already been carried out in a uniform
manner, employing the
$F^{\eta}-$test~\citep[e.g.,][]{Goyal2013MNRAS.435.1300G}. In this
test, it is specially important to use the correct rms errors on the
photometric data points. It has been found that the magnitude errors,
returned by the routines in the data reduction software DAOPHOT and
IRAF, are normally underestimated by a factor ranging between 1.3 and
1.75~\citep{Gopal-Krishna1995MNRAS.274..701G,
  Garcia1999MNRAS.309..803G, Sagar2004MNRAS.348..176S,
  Stalin2004JApA...25....1S, Bachev2005MNRAS.358..774B}.
Recently~\citep{Goyal2013JApA...34..273G} have estimated the best-fit
value of $\eta$ to be 1.54$\pm$0.05, using 262 sessions of intranight
monitoring of AGN.

The $F^{\eta}-$statistics can be written
as~\citep[e.g.,][]{Goyal2012A&A...544A..37G}

\begin{equation} 
 \label{eq.fetest}
F_{1}^{\eta} = \frac{\sigma^{2}_{(q-s1)}} { \eta^2 \langle
  \sigma_{q-s1}^2 \rangle}, \nonumber \\
\hspace{0.2cm} F_{2}^{\eta} = \frac{\sigma^{2}_{(q-s2)}} { \eta^2
  \langle \sigma_{q-s2}^2 \rangle},\nonumber \\
\hspace{0.2cm} F_{s1-s2}^{\eta} = \frac{\sigma^{2}_{(s1-s2)}} { \eta^2
  \langle \sigma_{s1-s2}^2 \rangle}
\end{equation}

where $\sigma^{2}_{(q-s1)}$, $\sigma^{2}_{(q-s2)}$ and
$\sigma^{2}_{(s1-s2)}$ are the variances of the `target-star1',
`target-star2' and `star1-star2' DLCs and $\langle \sigma_{q-s1}^2
\rangle=\sum_\mathbf{i=0}^{N}\sigma^2_{i,err}(q-s1)/N$, $\langle
\sigma_{q-s2}^2 \rangle$ and $\langle \sigma_{s1-s2}^2 \rangle$ are
the mean square (formal) rms errors of the individual data points in
the `target-star1', `target-star2' and `star1-star2' DLCs,
respectively. $\eta$ is the scaling factor and is taken to be $1.5$
(see  above).  \par

The $F$-values are calculated for each DLC using Eq.~\ref{eq.fetest}
and compared with the critical $F$ value,
$F^{(\alpha)}_{\nu_{qs},\nu_{ss}}$, where $\alpha$ is the significance
level set for the test, and $\nu_{qs}$ and $\nu_{ss}$ are the degrees
of freedom for the `target-star' and `star-star' DLCs (both are equal in the present work). Here, we set
two critical significance levels, $\alpha=$ 0.01 and 0.05, which
correspond to confidence levels of 99\% and 95\%, respectively. Thus,
we mark a NLSy1 as variable (‘V’) if F-value is found to be $>$
$F_{c}(0.99)$  for both its DLCs (relative to the two comparison stars), non-variable (‘NV’) if any one
out of two DLCs is found to have F-value $<$ $F_{c}(0.95)$. The
remaining cases are designated as probably variable (‘PV’).
The computed $F$-values and the corresponding INOV status for
the 3  $\gamma$-ray NLSy1s are given in columns 5 and 6 of
Table~\ref{NLSy1:tab_result}.\par

For computing the amplitude ($\psi$) of INOV we have followed the definition
given by~\citet{Heidt1996A&A...305...42H}

\begin{equation} 
\hspace{2.5cm} \psi= \sqrt{({D_{max}}-{D_{min}})^2-2\sigma^2}
\end{equation} 
with $D_{min,max}$ = minimum (maximum) values in the NLSy1-star DLC
and $\sigma^2 = \eta^2\langle\sigma^2_{q-s}\rangle$, where,
$\langle\sigma^2_{q-s}\rangle$ is the mean square (formal) rms error
of individual data point and $\eta$
=1.5~\citep{Goyal2013JApA...34..273G}.

\section{Results and Discussion}
\label{Sect 4}
It is worth reiterating that even though $\gamma$-ray loud NLSy1s
display blazar-like properties, their studies carry a special interest
because their AGNs reside in spiral galaxies (Sect.~\ref{Sect 1}) and thus the
jets are launched into a much denser environment than is the case for
blazars whose hosts are nearly always early-type galaxies ~\citep[see Sect.~\ref{Sect 1}; also,][and references therein]{Bagchi2014ApJ...788..174B}. The use of INOV as a tracer of blazar-like jet activity in $\gamma$-ray NLSy1s
became popular around the beginning of this decade. For the NLSy1 PMN
J0948$+$0022, ~\citet{Liu2010ApJ...715L.113L} observed a brightness
change of $\sim$ 0.5 mag over several
hours~\citep[also,][]{Eggen2013ApJ...773...85E,
  Paliya2013MNRAS.428.2450P, Liu2016NewA...44...51L}. Similarly,
violent INOV events have since been reported for a few other
$\gamma$-ray NLSy1s, such as
 1H 0323$+$342~\citep{Paliya2014ApJ...789..143P} and SBS 0846$+$513
(hereafter J0849$+$5108)~\citep[e.g.,][]{ Maune2014ApJ...794...93M,
  Paliya2016ApJ...819..121P}. It is now known that the duty cycle (DC)
of INOV for radio-loud NLSy1s is around $\sim$ 50\%, with somewhat
higher values found for their $\gamma$-ray detected
subset~\citep{Paliya2013MNRAS.428.2450P,
  Kshama2017MNRAS.466.2679K}. This, too, mirrors the situation known
for blazars~\citep[e.g.,][]{Stalin2005MNRAS.356..607S,
  Lister2009ApJ...696L..22L, Pushkarev2009A&A...507L..33P}. 
Below we summarize some salient aspects of the 3 $\gamma$-ray
NLSy1s as well as their variability properties found in the present study.

\subsection{The NLSy1 1H 0323$+$342 ($z$ = 0.0629)}

\label{Sect 4.1}
Multiple mass estimates for the central BH of this NLSy1 galaxy fall
in the range (1-3)$\times10^{7}
M_{\sun}$~\citep{Zhou2007ApJ...658L..13Z} which, although normal for
NLSy1s (Sect.~\ref{Sect 1}), is on the lower side for blazars
(Sect.~\ref{Sect 1}). Correspondingly, it is operating at a high Eddington ratio
of $\lambda_{Edd} \sim$ 0.4~\citep{Paliya2014ApJ...789..143P}, which
again is not exceptional for NLSy1s~\citep{Boroson2002ApJ...565...78B,
  Grupe2004ApJ...606L..41G}.  A blazar description of this NLSy1,
other than its $\gamma$-ray
flaring~\citep{Carpenter2013ATel.5344....1C,
  Paliya2014ApJ...789..143P}, stems from the flatness of its radio
spectrum up to 10 GHz~\citep[$\alpha_{r} =
  +0.1$,][]{Neumann1994A&AS..106..303N} and even higher radio
frequencies~\citep{Angelakis2015A&A...575A..55A}. Another evidence for
a relativistically beamed jet comes from its VLBI image, which shows a
radio core with one-sided
jet~\citep[e.g.,][]{Lister2005AJ....130.1389L,
  Zhou2007ApJ...658L..13Z}. In their detailed VLBI study of this source at 15 GHz,~\citet{Fuhrmann2016RAA....16..176F} have resolved the jet into 7 well-aligned knots and estimated them to have proper motions of up to 7c. They also estimate the jet’s viewing angle to be within $\sim$ 13 degrees. Furthermore, in the states of both
average and high $\gamma$-ray activity, its dual-humped SED showed the
synchrotron component peaking near 10$^{12.5}$ Hz, which is a
characteristic of FSRQ/LBL type AGN~\citep{Zhou2007ApJ...658L..13Z,
  Paliya2014ApJ...789..143P}.

By comparing the SED for its nucleus with other
AGN,~\citet{Zhou2007ApJ...658L..13Z} concluded that its optical light
is dominated by thermal emission, as also independently inferred
by~\citet{Paliya2014ApJ...789..143P} from their SED modeling for both
the average and high-activity states. The contamination from this
thermal emission of Seyfert origin is probably responsible for the observed low
optical polarization $<$ 1\% ~\citep{Eggen2011nlsg.confE..49E,
  Ikejiri2011PASJ...63..639I}.  Even during the high activity state
detected by {\it Fermi}/LAT in July 2013, which lasted $\sim$ 20 days,
its polarization rose only to $\sim$
3\%~\citep{Itoh2014PASJ...66..108I}. A similarly low polarisation has been reported by~\citet{Pavlidou2014MNRAS.442.1693P} and, more recently by~\citet{Angelakis2018arXiv180702382A}. At radio wavelengths
too, the source exhibited a rather modest polarisation~\citep[p$_{rad}
  \sim$ 4\% at 10.55 GHz,][]{Neumann1994A&AS..106..303N}. Recall, however, that 
polarisation dips are not unexpected for NLSy1
galaxies~\citep{Eggen2011nlsg.confE..49E, Ikejiri2011PASJ...63..639I,
  Itoh2013ApJ...775L..26I, Itoh2014PASJ...66..108I}.
Even for blazars,~\citet{Fugmann1988A&A...205...86F} has shown that
there is $\sim$ 40\% chance that a {\it bona fide} blazar will appear
only weakly polarised (p$_{opt} <$ 3\%) at a random epoch.  Recall also
that even the very prominent BL Lac object OJ 287 has sometimes been
found to be unpolarized~\citep[e.g.,][]{Villforth2009MNRAS.397.1893V}.\par

Taking only the unbeamed radio flux, ~\citet{Zhou2007ApJ...658L..13Z}
found an R$_{5 GHz}$ between 4 and 25, placing this NLSy1 in the category of `radio
intermediate quasars' (RIQs)~\citep{Miller1993MNRAS.263..425M,
  Falcke1996ApJ...473L..13F, Falcke1996ApJ...471..106F,
  Diamond-Stanic2009ApJ...698..623D}.  Here we may note the extensive
dataset on intranight optical monitoring of RIQs, published
by~\citet{Goyal2010MNRAS.401.2622G}, has demonstrated that their INOV
generally maintains a low level, both in amplitude ($\psi <$ 3\%) and
duty cycle (DC $\sim$ 10\%). Putting together this and the modest
polarisation, it would seem unlikely that a strong INOV activity can
be witnessed in this NLSy1. However, as we discuss below, this
somewhat discouraging prognosis was overturned by the two episodes of
violent optical variability recorded for this NLSy1 in separate
intranight monitoring sessions $\sim$ 4 years apart.\par

The first INOV study of this $\gamma$-ray NLSy1 was reported
by~\citet{Paliya2013MNRAS.428.2450P} who monitored it on 4 nights
within a span of 10 days during early 2012. The observations on two of
the nights were quite noisy and in the remaining two nights, the
monitoring duration exceeded our threshold of 3 hours just on the
night of 26.01.2012, when the source showed a strong INOV ($\psi \sim$
7\%). In another campaign during late 2012, the source was monitored
by them on 3 nights~\citep{Paliya2014ApJ...789..143P}. On the first
two nights, only mild INOV was observed, however, on the third night
(09.12.2012) the DLCs taken with a temporal resolution of 2 minutes,
showed a very strong outburst, when the optical flux rose by $\psi \sim 35$\% within just $\sim$ 30 minutes. The import of this
and a similar INOV event detected in the present study of this NLSy1
is discussed below.\par
In our campaign during 2016-17, we monitored this
NLSy1 on 5 nights, each time for $\geq$ 3 hours
(Table~\ref{NLSy1:tab_result}).  Strong INOV, with $\psi$ between
  4\% and 9\% was detected on 4 of the 5 nights
(Table~\ref{NLSy1:tab_result}; Fig.~\ref{fig:lurve 1}~\&~\ref{fig:lurve 2}, see below),
which is clearly reminiscent of blazars~\citep[e.g.,][]{Goyal2013MNRAS.435.1300G}. In all these sessions,
seeing disc variations were quite small; the seeing data for the night
of 02.12.2016 are plotted in Fig.~\ref{fig:lurve 3}. It may be
recalled that the issue of variable seeing is specially relevant for
nearby AGN like this NLSy1, for which the host galaxy can be a
significant contributor to the aperture photometric measurements
~\citep{Cellone2000AJ....119.1534C}. Fortunately, an HST
image is available for this NLSy1, and it shows that the host galaxy,
with a total extent of 15 arcsec, contributes close to 50\% of the
optical flux, the remainder coming from the
AGN~\citep{Zhou2007ApJ...658L..13Z}. As discussed below, this
information plays a key role in a quantitative interpretation of the
INOV and other observations of this AGN.\par

To further check the possible impact of seeing variations on our DLCs,
we have derived a new set of DLCs taking larger
photometric apertures (i.e., radius = 4 , 6 and 7 times the median
FWHM found for the given session). These DLCs confirm the strong INOV seen
in the DLCs of this source (see Fig.~\ref{fig:lurve 1}), for all the nights, excepting the night of 
03.01.2017 (Table~\ref{NLSy1:tab_result}). Focussing next on the session on 02.12.2016 when a large optical outburst was seen, Fig.~\ref{fig:lurve 3} shows that the base level of the `AGN - star' DLCs stops rising further significantly when the aperture radius
crosses 6$\times$FWHM (15.6 arcsec). This means that this aperture is
large enough to pick virtually the entire emission from the host
galaxy (which is consistent with its size in the HST image taken
by~\citet{Zhou2007ApJ...658L..13Z}.\\ To summarize, the present INOV
observations have confirmed that this NLSy1 galaxy is capable of
strong INOV activity, with a high DC $\sim$ 60 to 75 per cent. Earlier, a
similarly intense INOV activity has been observed for the
proto-typical NLSy1s
J0849$+$5108~\citep{Maune2014ApJ...794...93M}. Such large and
frequent INOV is strikingly reminiscent of some prominent BL Lacs,
like AO 0235+164 and OJ 287~\citep[e.g.,][]{Romero2000A&A...360L..47R,
  Sagar2004MNRAS.348..176S, Goyal2017arXiv170904457G,
  Britzen2018MNRAS.478.3199B}.

\subsection{The spectacular optical outburst of 1H 0323$+$342}
\label{Sect 4.2}
  Although INOV was detected for the NLSy1 galaxy 1H 0323$+$342 on
  4 of the 5 nights we monitored it, the most spectacular variation
  occurred on 02.12.2016. During the 4.4 hours of continuous
  monitoring with high-sensitivity, both comparison stars remained
  rock steady and the seeing disk, too, was steady
  (Fig.~\ref{fig:lurve 3}). Nearly in the middle of the session, a
  large, roughly flat-topped and nearly symmetric outburst of total duration
  $\sim$ 1.25 hours was observed. For the DLCs corresponding to the
  aperture radius of 6$\times$FWHM (see above), the rising phase of the outburst,
  which is temporally resolved, shows a 0.07 mag variation occurring
  within at most 20 minutes and a similarly steep gradient was seen for the
  declining phase which, too, is resolved temporally. Such sharp
  variations are extremely rare episodes even for blazars
  ~\citep[e.g.,][and references therein]{Gopal-Krishna2018BSRSL..87..281G}. Curiously, this
  outburst bears an uncanny resemblance to the one this NLSy1 had
  exhibited on 09.12.2012~\citep[Fig. 10
    of][]{Paliya2014ApJ...789..143P} coinciding with a $\gamma$-ray
  flare (note that the precursor optical bump seen in the DLCs on that
  night is most probably an artefact due to the sudden spell of seeing disk
  deterioration, which can be seen in the bottom panel of
  their Fig. 10). During that outburst, this NLSy1 brightened by
  $\sim$ 0.35 mag in 30 minutes and after remaining at the elevated
  brightness for $\sim$ 1.1 hours, reverted almost as rapidly to its
  initial level. While the amplitudes of these two optical outbursts
  are impressive, they are by no means exceptional for $\gamma$-ray
  NLSy1s. For instance, two INOV flares of $\psi \sim$ 0.3 mag, with a
  rise/fall time between 10-30 minutes were detected during the
  intranight monitoring of the NLSy1 PMN J0948$+$0022 on
  01.04.2011~\citep{Eggen2013ApJ...773...85E}. Earlier,~\citet{Liu2010ApJ...715L.113L}
  had reported for the same NLSy1 a brightness change of $\sim$0.5 mag
  over several hours. Likewise, during a high $\gamma$-ray activity phase, the
  NLSy1 J0849$+$5108 was found to fade by $\sim$ 0.2 mag
  within just $\sim$ 15 minutes~\citep[figure 6
    of][]{Maune2014ApJ...794...93M}. As we shall now argue for
  1H 0323$+$342, the 0.07 mag brightening within {\it at most} 20
  minutes, seen at the beginning of the 02.12.2016 outburst
  (Fig.~\ref{fig:lurve 2}~\&~\ref{fig:lurve 3}), actually corresponds to a remarkably short
  flux doubling time of $\sim$ 1 hour for this AGN's
  {\it nonthermal} output. This conclusion is reached when we subtract out from the aperture photometric measurements, the expected contributions of thermal optical emission, made by the host galaxy and by the accretion disk associated with the Seyfert nucleus~\citep [even which is not expected to vary by $>$ 1\% on hour-like time scale; see][]{Mangalam1993ApJ...406..420M}. To get an
  idea of the accretion disc's contribution, we return to the modeling
  of this AGN's SEDs for the four epochs, which revealed that in both
  high and low states of $\gamma$-ray activity, the optical emission
  was dominated by the thermal component contributed by its Seyfert
  nucleus~\citep[][their Fig. 9]{Paliya2014ApJ...789..143P}, in
  agreement with the conclusion reached earlier
  by~\citet{Zhou2007ApJ...658L..13Z}. Accordingly, we shall make a
  conservative assumption that 50\% of the optical emission from the
  AGN is thermal and only the remainder is synchrotron light.  Next,
  consider the HST image of this NLSy1 which showed that the total
  optical emission from the (unresolved) AGN is essentially equal to
  that arising from the underlying host galaxy $\sim$ 15 arcsec in
  diameter~\citep{Zhou2007ApJ...658L..13Z}. As discussed above,
  essentially all this emission from the host galaxy has got picked up
  in our photometry with a circular aperture of radius 6$\times$FWHM
  $\sim$ 15.6 arcsec. Thus, putting together the likely thermal
  contributions to the DLCs of this NLSy1 (Fig.~\ref{fig:lurve 3}),
  which come from the Seyfert nucleus and the host galaxy, it can be
  concluded that only $\sim$ 25\% or less of the amplitude of the
  light curves is of synchrotron origin and the observed rapid
  outburst and other observed INOV is associated entirely with this
  minor component. Therefore, in order to account for the observed
  brightening by $\sim$ 7\%  at the beginning of the optical outburst, the
  optical synchrotron component of the AGN is required to have
  brightened up (in $<$ 20 minutes, see above) by a factor of
  1.27. This corresponds to a flux doubling time of $\sim$ 1
  hour. Such INOV can only be described as extreme, as it is about 20 times larger in amplitude than that typically displayed by blazars ~\citep[e.g.,][]{Ferrara2001ASPC..224..319F, Goyal2012A&A...544A..37G}. If, in a less conservative vain, we
  consider the share of synchrotron optical emission in the total
  optical output of the AGN to be less than 50\%~\citep[see,
    e.g.,][]{Paliya2014ApJ...789..143P}, the deduced flux doubling
  time would be shorter still.\par

Note that a very similar conclusion can be drawn from the optical
flare exhibited by this NLSy1 on 9.12.2012~\citep[Fig. 10
  of][]{Paliya2014ApJ...789..143P}.  During that flare, the source
brightened up by $\sim$ 35\% in just 30 minutes. As opposed
to our DLCs displayed in Fig.~\ref{fig:lurve 2}~\&~\ref{fig:lurve 3}, those DLCs are based
on photometry with a much smaller aperture; indeed, they are claimed
by~\citet{Paliya2014ApJ...789..143P} to essentially represent just the
AGN component of the emission, which they assert is dominated by thermal
radiation. Before proceeding further, we make two conservative
assumptions about those DLCs of J0324+3410, namely that (a) they have zero
contribution from the host galaxy and (b) that 50\% (i.e., the maximum
permissible) of the amplitude of each DLC is due to the AGN's
synchrotron radiation. Even on this grossly conservative basis, it is
evident that, in order to cause the observed brightness change of $\sim $ 35\% in 30
minutes, the AGN's {\it synchrotron} optical emission must have
increased by at least $\sim$ 70\% (within the 30 minutes). This
corresponds to a flux doubling time of $\sim$ 0.7 hours, a yet another
exceptional event exemplifying once again the extreme INOV behaviour
of this NLSy1. In reality, the flux doubling times for both this flare
and the one reported here (Fig.~\ref{fig:lurve 2}~\&~\ref{fig:lurve 3}) may be
substantially shorter than the present conservative estimates of
$\leq$ 1 hour, thereby approaching the extremely fast variability of
GeV/TeV radiation detected for some blazars~\citep[e.g., a flux
  doubling time of $\sim$ 10 minutes observed for the blazar PKS
  1222+21 at 400 GeV, by][]{2011ApJ...730L...8A,
  Ackermann2014ApJ...786..157A}.

\subsection{The NLSy1 PKS J1222$+$0413 ($z$ = 0.966)} 
\label{Sect 4.3}
This is the farthest known $\gamma$-ray-emitting NLSy1. Aside from its
30$\sigma$ level of $\gamma$-ray detection~\citep[][hereafter
  YOF15]{Ackermann2015ApJ...810...14A, Yao2015MNRAS.454L..16Y} and the
VLBI detection of a one-sided jet~\citep{Lister2016AJ....152...12L},
evidence for relativistic jet in this NLSy1 comes also from its
extreme radio loudness (R$_{5GHz} \sim$ 1700 for the core, YOF15), a
relatively flat hard X-ray spectrum (photon index $\Gamma \sim$ 1.3,
YOF15) which is typical of inverse Compton X-rays from relativistic
jets, and an inverted radio spectrum~\citep[$\alpha_{r}$ = +0.3
  between 1.4 and 4.9 GHz,][]{White1992ApJS...79..331W}. Others
evidence include a high brightness temperature of its VLBI
core~\citep[$\sim 4\times10^{12}$ K at 8.6
  GHz,][]{Pushkarev2012A&A...544A..34P} and a rather large long-term
radio flux variability, ranging from $\sim$ 0.5 to 1.1 Jy at 5 GHz
(YOF15). The synchrotron bump in its bi-modal SED peaks in the
infrared (YOF15), a hallmark of LBL type blazars
~\citep{Urry1995PASP..107..803U, Abdo2010ApJ...716..835A} which are
known to exhibit the strongest optical variability among all AGN
classes, both on hour-like and longer time scales~\citep[e.g.,][]
{Heidt1996A&A...305...42H, Hovatta2014MNRAS.439..690H}.

Modeling the SED in terms of one-zone leptonic relativistic jet
undergoing external Compton losses has yielded a bulk Lorenz factor
$\Gamma \sim 30 \pm$ 5 (YOF15) which is very close to the similarly
estimated $\Gamma \sim$ 30 for the well-known radio-loud $\gamma$-ray
NLSy1 PMN J0948$+$0022~\citep{D'Ammando2015MNRAS.446.2456D}, but much
larger than $\Gamma <$ 15 which is typical for $\gamma$-ray
NLSy1s~\citep[e.g.,][]{Abdo2009ApJ...707..727A,
  D'Ammando2012MNRAS.426..317D}. Interestingly, a similar excess
exists even in the BH mass. The best virial estimates, based on
optical spectra, are (1-2)$\times10^{8}M_{\sun}$ for
PKS J1222+0413~\citep[YOF15,][]{Sbarrato2012MNRAS.421.1764S} and
$\sim10^{8}M_{\sun}$ for the archetypal $\gamma$-ray NLSy1 PMN
J0948$+$0022~\citep{Zhou2003ApJ...584..147Z}. These are much higher
than the typical value (M$_{BH} \sim 10^{7} M_{\sun}$) reported for
NLSy1s~\citep[e.g.,][]{Yuan2008ApJ...685..801Y} although these
estimates could be systematically low for reasons mentioned in
Sect.~\ref{Sect 1}~\citep[e.g., see][]{Calderone2013MNRAS.431..210C}, albeit
contested by others~\citep[e.g.,][]{Jarvela2017A&A...606A...9J}. In
spite of the atypically high M$_{BH}$, the central engine of
PKS J1222+0413 is found to operate at a very high Eddington ratio (
$\lambda_{Edd} \sim$ 0.6, YOF15).

To our knowledge, the present study is the first characterisation of
the INOV of this NLSy1. Out of our 5 nights of monitoring, during
January-March (2017), INOV was seen in just one session (22.02.2017)
when the source showed a steady brightening by $\sim$ 0.1 mag over 3
hours (Fig.~\ref{fig:lurve 1}). Excluding this session with a clear
INOV detection, the maximum brightness change witnessed across the
remaining 4 sessions is only $\sim$ 0.08 mag, which is not excessive
even for radio-quiet quasars of low-luminosity/redshift
~\citep[e.g.,][]{Caplar2017ApJ...834..111C}. Note, however, that the
3.4 and 4.6 micron data in the WISE
catalogue~\citep{Wright2010AJ....140.1868W} have shown an intra-day
variation with $\psi \sim$ 50\% (YOF15), which is clearly blazar-like
and a similar large amplitude has been witnessed in the radio band
where its 5 GHz emission was found to vary in the range from $\sim$
0.5 to 1.1 Jy over a time scale of years (YOF15).

\subsection{The NLSy1 PKS J1505$+$0326 ($z$ =0.4089) }
\label{Sect 4.4}

The {\it Fermi}/LAT detection of this `extremely radio-loud'
(R$_{1.4GHz} \sim$ 3364) NLSy1~\citep{Abdo2009ApJ...707L.142A}
constitutes a strong evidence for a relativistically boosted
nonthermal jet. Additional evidences include the relatively hard X-ray
spectrum~\citep{D'Ammando2013MNRAS.433..952D}, an inverted radio
spectrum up to $\sim$ 10 GHz ($\alpha_{r}\sim+$0.3) and a high
brightness temperature inferred from flux variability at 2.6 GHz,
$\sim 2.6\times10^{13}$
K~\citep{Angelakis2015A&A...575A..55A}. Another indication of a high
Doppler factor (between 3.9 and 6.6) has come from radio flux
variability at 15 GHz during the $\gamma$-ray
outbursts~\citep[][hereafter DOF16]{D'Ammando2013MNRAS.433..952D,
  D'Ammando2016MNRAS.463.4469D}, invoking the concept of
`equipartition Doppler Factor'~\citep{Singal1985MNRAS.215..383S,
  Readhead1996ApJ...460..634R}.
From VLBI, the source is also known to exhibit a bright core plus a
weak jet at 15 GHz~\citep{D'Ammando2013MNRAS.433..952D,
  Orienti2012arXiv1205.0402O}. However, the VLBI images collected
during 2010 - 2013 under the MOJAVE\footnote{Monitoring Of Jets in
  Active galactic nuclei with VLBA Experiments
  (https://www.physics.purdue.edu/MOJAVE/)}
programme~\citep{Lister2016AJ....152...12L} have revealed only a
sub-luminal component in its radio jet, with a proper motion of
1.1c$\pm$0.4 c.  The only major $\gamma$-ray flaring activity of this
NLSy1, recorded by {\it Fermi}/LAT, occurred during Dec. 2015 to
Jan. 2016~\citep{D'Ammando2015ATel.8447....1D} when its light-curve
plotted in 3-hour bins showed a brightening by almost 2 orders of
magnitude, although no significant signal was detected on shorter
timescale.  Multi-band observations triggered by this rare flaring
event have been reported by~\citet{Paliya2016ApJ...820...52P} and
DOF16. The SEDs during the flare activity and the state of average
activity, both show the synchrotron bump peaking near 10$^{13.5}$ Hz,
which is characteristic of FSRQs/LBLs and unlike HBLs. This conclusion
is reinforced by its observed similarity to LBLs in terms of
$\gamma$-ray luminosity, photon index and Compton
dominance~\citep{Ackermann2015ApJ...810...14A,
  Paliya2016ApJ...820...52P, D'Ammando2016MNRAS.463.4469D}.

From modeling the SED's optical/UV bump as Shakura-Sunyaev accretion
disk, $M_{BH} \sim 4\times10^{7} M_{\sun}$ has been estimated
by~\citet{Paliya2016ApJ...820...52P}, which is $\sim$ 2 times the
virial estimate reported by~\citet{Shaw2012ApJ...748...49S} based on
the Mg II line.  Taking this M$_{BH}$ and modeling the SED in terms of
the standard single leptonic blob moving in a relativistic jet close
to our direction and producing synchrotron self-Compton (SSC)
and external Compton (EC)
emissions~\citep[e.g.,][]{Tavecchio2001ApJ...554..725T,
  Finke2008ApJ...686..181F, Dermer2009ApJ...692...32D,
  Abdo2011ApJ...736..131A}, yields $\Gamma_{jet} \sim$ 17 during the
high activity state, which corresponds to an Eddington ratio of $\sim$
0.1 (DOF16).\par In the optical band, this NLSy1 has been under
regular monitoring since 2008 in the
CRTS~\citep{Drake2009ApJ...696..870D}. Until 2016 a total variability
amplitude of 0.7 mag has thus been recorded (DOF16), which is
blazar-like ~\citep[see, e.g.,][]{2014A&A...569A..95K}. On the
intranight scale, prior to our
observations,~\citet{Paliya2013MNRAS.428.2450P} had monitored this
NLSy1 on 4 nights during April-May (2012) which coincided with its
moderately active $\gamma$-ray phase (DOF16). While the resulting
optical DLCs are not available, making it hard to assess their
sensitivity level and cadence, the authors have reported a clear INOV
detection with $\psi \sim$ 10\% on one of the 4 nights
(24.05.2012). This amplitude too, is blazar-like~\citep[see,][]{
  Ruan2012ApJ...760...51R, Goyal2013MNRAS.435.1300G,
  2014A&A...569A..95K}.  ~\citet{Paliya2013MNRAS.428.2450P} also
detected an episode of (mild) inter-day optical variability of $\sim$
0.045 mag and this too is on the higher side for radio-quiet quasars,
which are known to vary typically by atmost 1-2\% on 1-day time
scale~\citep[e.g.,][]{Caplar2017ApJ...834..111C}.  The results of our
monitoring of this extremely radio-loud NLSy1 on 5 nights are
displayed in Fig.~\ref{fig:lurve 1}. The only significant variability
is the $\sim$ 0.1 mag brightness change between 21.04.2018 and
11.05.2018, representing short-term optical variability (STOV) on 
day-like time scale.

In summary, it seems fair to conclude that during our 5 nights'
optical monitoring, the several previously known blazar-like
attributes of this extremely radio-loud NLSy1 (Sect.~\ref{Sect 4.1}) were not robustly
 reflected in its INOV and longer-term optical variability, which stands in
stark contrast to the NLSy1 1H 0323$+$342, the least radio-loud member
of the present set of 3 $\gamma$-ray NLSy1s. As a caveat, we would
like to mention that the much lower level of INOV shown by
PKS J1505$+$0326 may have to be revised upwards if in future it becomes
possible to reliably account for the dilution by the thermal optical
output of its AGN, for which a hint already
exists~\citep{D'Ammando2016Galax...4...11D}.

\subsection{Comments on the INOV diversity of the 3 $\gamma$ -ray NLSy1s} 
\label{Sect 4.5}

The 3 {\it Fermi}/LAT detected NLSy1 galaxies
(Table~\ref{tab:source_info}) studied here, share a number of
observational commonalities, like highly significant $\gamma$-ray and
radio detection, a flat/inverted radio spectrum (which flags dominance
by a parsec-scale, or more compact radio jet). Additionally, their
broadband SEDs are not only blazar-like (with a dual-hump profile) but
also the peaking of the synchrotron hump in the infrared in each case,
means that all 3 NLSy1s are counterparts of the FSRQ/LBL subclass of
blazars and hence expected to exhibit strong radio/optical flux
variability~\citep[e.g.,][]{ Heidt1997A&A...325...27H,
  Maune2014ApJ...794...93M}. In the present study, a blazar-like
intense and frequent INOV (duty cycle $\sim$ 60 - 75\%) was detected
only for the NLSy1 1H 0323$+$342. Remarkably, not only is this NLSy1 the least radio-loud
among the 3 NLSy1s reported here, a recent study has even re-classified it
as a `radio intermediate' with R$_{5 GHz}$ between 4 and 25, based on its
unboosted radio emission~\citep{Zhou2007ApJ...658L..13Z}. Even starker
 manifestations of its violent INOV are encapsulated in the two
optical outbursts for which the present analysis conservatively
implies a doubling time of $\sim$ 1 hour for the optical synchrotron flux
(Sect.~\ref{Sect 4.2}). Such an INOV activity is intriguing for a `radio
intermediate' since an extensive study has shown that, in general,
radio-intermediate quasars exhibit only low-level INOV ($\psi <$
3\%) and the duty cycle is also small~\citep[$\sim$ 10\%,
  see][]{Goyal2010MNRAS.401.2622G}. The contrary behaviour of the
NLSy1 J0324$+$3410 indicates that a large radio loudness parameter
{\it per se} is not an essential pre-requisite for strong INOV in the
case of NLSy1 galaxies. Even for radio-loud quasars, a high radio
loudness does not guarantee a strong INOV, as emphasized in
\citet{Goyal2012A&A...544A..37G},~\citep[see
  also,][]{Gopal-Krishna2018BSRSL..87..281G}. Specially in the
context of NLSy1 galaxies, a closer look into the dependence of INOV
on radio-loudness would need a more realistic estimation of
radio-loudness, since the required exclusion of the Doppler boosted
anisotropic radio emission can make a big difference, given the low
prominence of diffuse radio emission in NLSy1s with prominent
jets~\citep{Doi2012ApJ...760...41D, Congiu2017A&A...603A..32C}.\par
For $\gamma$-ray NLSy1s, uncertainty in R$_{5GHz}$ is not the only
complicating factor in probing the dependence of INOV on radio
loudness. As highlighted above from the example of 1H 0323$+$342, the
observed INOV amplitude may itself be significantly, if not grossly,
underestimated, depending on the extent to which the AGN's rapidly varying synchrotron
optical luminosity (arising from its relativistic jet) gets
contamination by the AGN's thermal light. Even for the radio detected
NLSy1s, there is growing evidence for a significant, if not dominant
thermal component in the AGN's optical
output~\citep[e.g.,][]{D'Ammando2012MNRAS.426..317D,
  Sbarrato2018arXiv180506459S}, even in an active
state~\citep[e.g.,][]{Zhou2007ApJ...658L..13Z,
  Paliya2014ApJ...789..143P}. Another potentially serious
observational repercussion of such a thermal contamination is the
dilution of the AGN's optical polarization signal, as emphasized in
Sect.~\ref{Sect 4.1} for the specific case of the NLSy1 1H 0323$+$342. Such a
camouflaging of the blazar lurking inside the nucleus is also known to
occur in some quasars~\citep[e.g.,][]{Giroletti2009ApJ...706L.260G,
  Antonucci2012A&AT...27..557A}, 3C273 being a prime
example~\citep{Impey1989ApJ...347...96I, Valtaoja1990AJ.....99..769V,
  Courvoisier1998A&ARv...9....1C}.  But, this problem is probably more prevalent
 in the case of radio-loud NLSy1s, since they are generally
high Eddington-rate
accretors~\citep[e.g.,][]{Foschini2015A&A...575A..13F}, hence expected
to generate a larger thermal radiation in the Seyfert
nucleus~\citep[besides intermittent jet activity,
  see,][]{Czerny2009ApJ...698..840C}. Thus, at least for defining the
polarisation status, radio band may offer a more fruitful channel, since
any thermal contamination by the AGN is expected to be far less
substantial, although perhaps not vanishingly small~\citep[see,
  e.g.,][]{Laor2008MNRAS.390..847L}. Another challenge is to arrive at
a proper estimate of radio-loudness (i.e., R$_{5 GHz}$) for which knowledge of
the unbeamed radio emission is vital. Mostly, this would require sensitive
VLBI observations, given that the radio spectrum associated with
$\gamma$-ray NLSy1s is usually flat or
inverted~\citep[e.g.,][]{Angelakis2015A&A...575A..55A}, which means
that the unbeamed radio emission is either faint, or compact (or,
both). All this underscores the increasing role of radio observations
in probing the physics of NLSy1s.\par
Nevertheless, one point that stands out from
Table~\ref{tab:source_info}, is that the NLSy1 1H 0323$+$342, the
solitary AGN showing a large and frequent INOV in our program thus
far, is also the only one for which VLBI monitoring has revealed a
large apparent superluminal motion in the jet (v $\sim$ 9c); the other two
NLSy1s (PKS J1222$+$0413 and PKS J1505$+$0326) have only displayed a
sub-luminal synchrotron jet (of radio/optical emission, see
Table~\ref{tab:source_info}). Here, it needs to be clarified that
although the bulk Lorenz factors for these two sources, as determined
from SED modeling are actually very large ($\Gamma \sim$ 30 and $\sim$
17), these values are more representative of the parts of the jet
which produce high-energy photons and probably not closely linked to
the radio/optical emitting zones of the relativistic
jet~\citep[e.g.,][]{ Ghisellini2005A&A...432..401G}. It is also
interesting to consider the other two well known {\it Fermi}/LAT
NLSy1s, namely J0849$+$5108 and PMN J0948$+$0022, both of which have
exhibited intense INOV~\citep[e.g.,][]{Eggen2013ApJ...773...85E,
  Maune2014ApJ...794...93M}.  Their VLBI monitoring at 15 GHz under
the MOJAVE programme has demonstrated that both are superluminal, with
an apparent speed of 5.8c$\pm$0.9c for J0849$+$5108, and 11.5c$\pm$1.5c
for PMN J0948$+$0022~\citep{Lister2016AJ....152...12L}. Thus, at
present, no example of $\gamma$-ray emitting NLSy1s is known where a
strong INOV is associated with a radio jet lacking apparent
superluminal motion. It was noted above that the persistent violent
INOV activity of 1H 0323$+$342 is mirrored in the BL Lac object
AO0235$+$164. It is interesting that this blazar has displayed an
ultra-fast apparent superluminal speed of $\sim$
46c~\citep{Jorstad2001ApJS..134..181J}.\par
In summary, the present study of 3 {\it Fermi}/LAT NLSy1 galaxies is
suggestive of a physical link between violent INOV and apparent
superluminal motion, two key observables for active galactic
nuclei. This is not to contend that the INOV relates weakly to other
important parameters, e.g., optical polarization, which would be in
clear conflict with the strong correlation found in the case of
radio-loud quasars~\citep[e.g.,][]{Goyal2012A&A...544A..37G}. But, as
mentioned above, verifying such a linkage in the case of NLSy1s would
require a more concerted radio/optical follow-up that would enable a
fairly precise separation of the synchrotron and thermal radiations in
the SED of the Seyfert nucleus. The challenges in realising this make
NLSy1 galaxies important targets for multi-wavelength, time-domain astronomy.

\begin{figure*}
\centering
\epsfig{figure=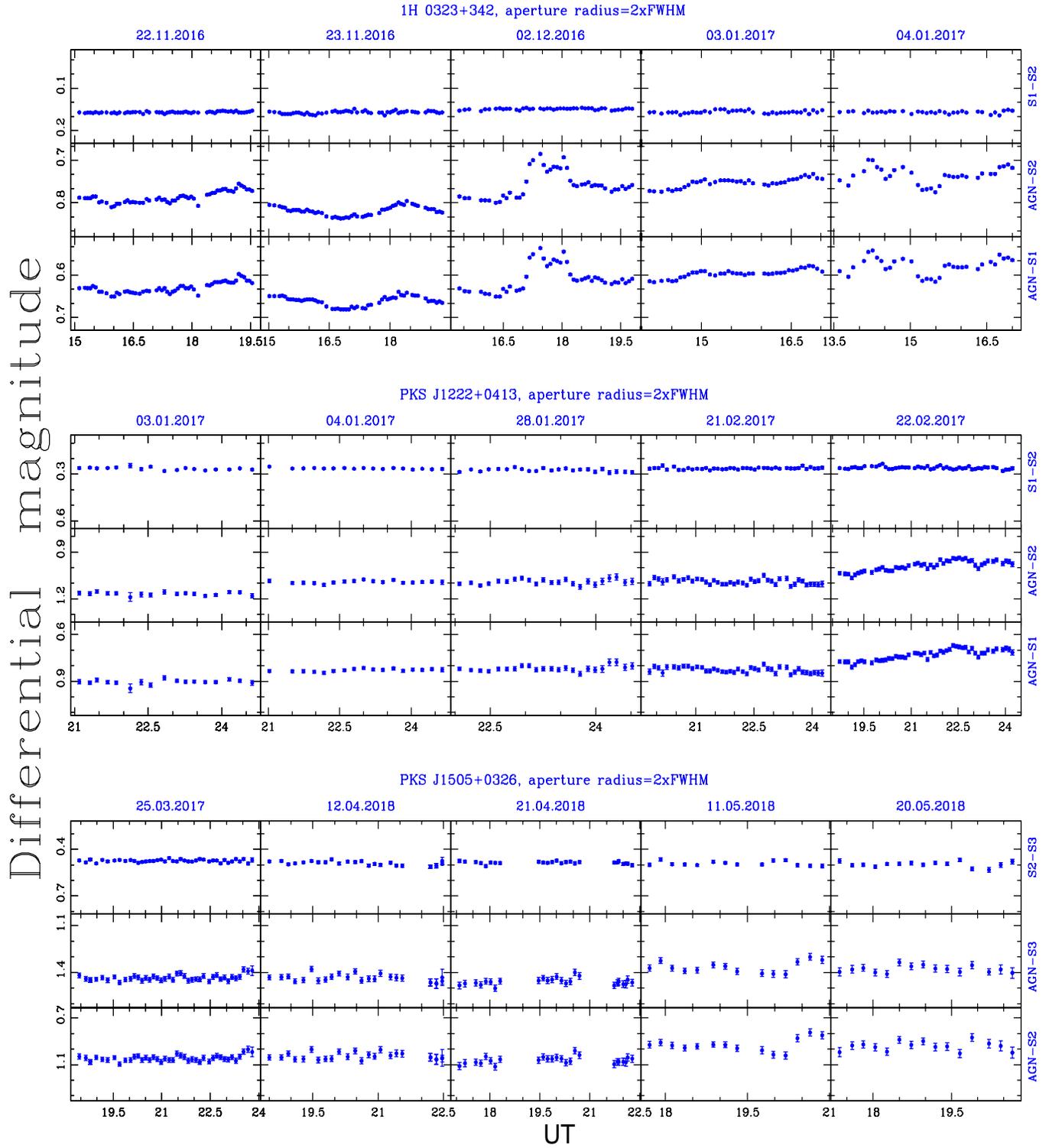,height=20.0cm,width=18.0cm,angle=0,bbllx=18bp,bblly=143bp,bburx=590bp,bbury=717bp,clip=true}
\caption[]{Medium/long - term DLCs of the three $\gamma$-ray NLSy1s.
  The names of the NLSy1s and their dates of observations are given at the top of the panels. In
  each panel, the upper DLC is derived using the two non-varying
  comparison stars, while the lower two DLCs are the `NLSy1-star'
  DLCs, as defined in the labels on the right side.} 
\label{fig:lurve 1}
\end{figure*}

\begin{figure*}
\centering
\epsfig{figure=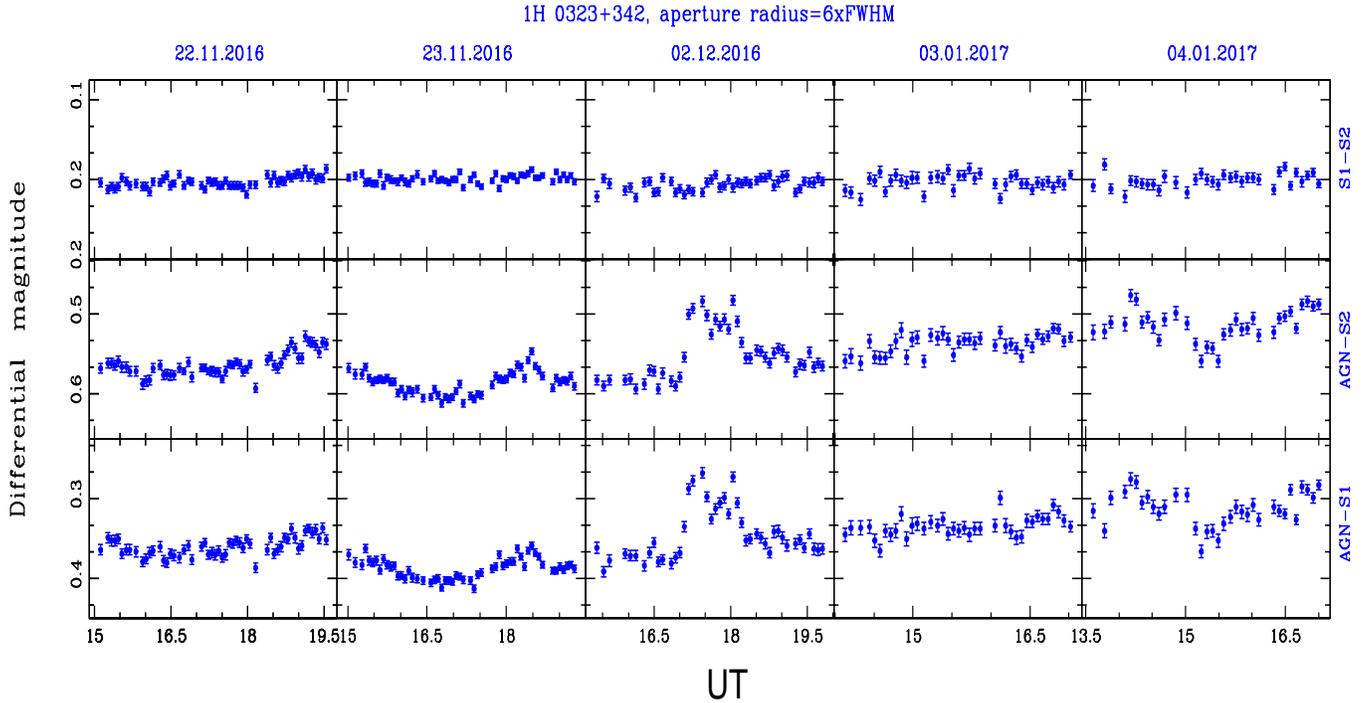,height=9.5cm,width=18.0cm,angle=0,bbllx=18bp,bblly=483bp,bburx=591bp,bbury=715bp,clip=true}
\caption[]{Same as Fig.~\ref{fig:lurve 1} for 1H 0323$+$342, but for an aperture radius = 6$\times$FWHM.}
\label{fig:lurve 2}
\end{figure*}

\begin{figure}
\epsfig{figure=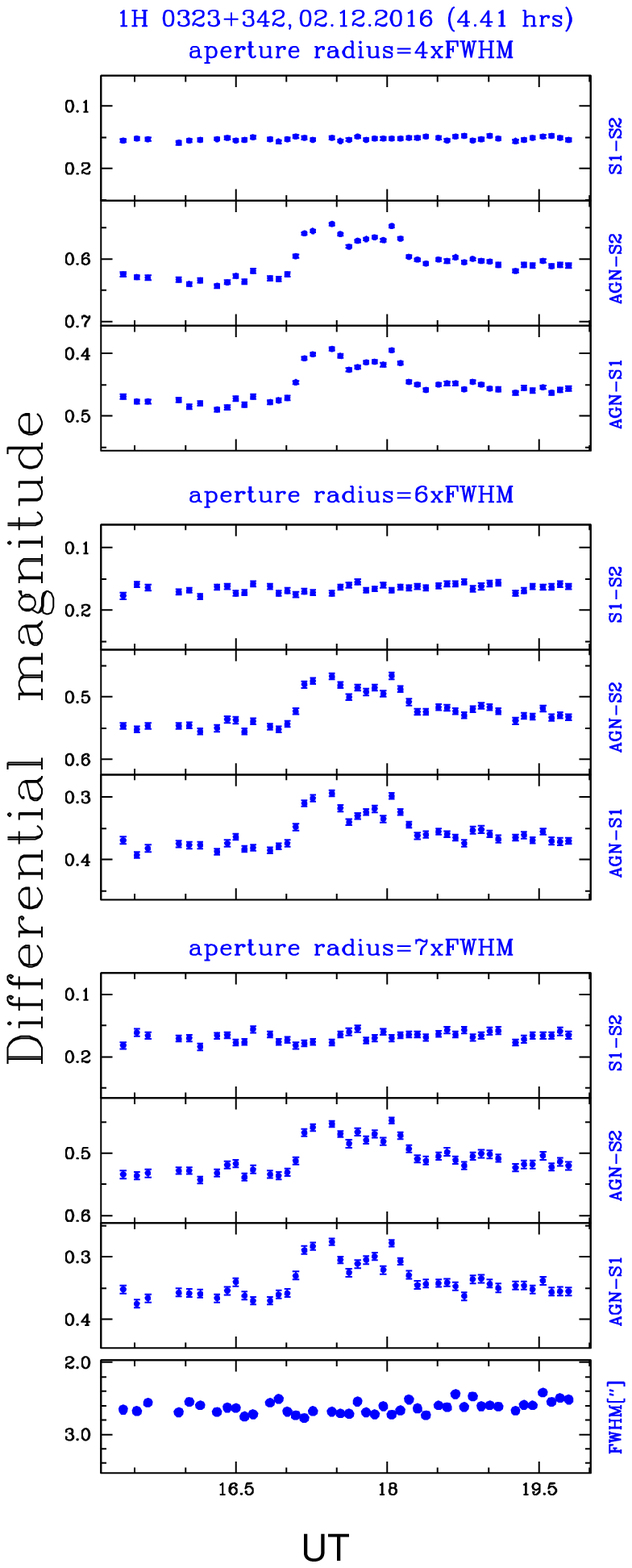,height=22cm,width=8.5cm,angle=0,bbllx=12bp,bblly=145bp,bburx=242bp,bbury=712bp,clip=true}
\caption[]{DLCs of 1H 0323$+$342 derived using three aperture radii, 4$\times$FWHM, 6$\times$FWHM and 7$\times$FWHM and plotted in the figure from the top to bottom panels, respectively, along with the seeing profile for the session (bottom panel).}
\label{fig:lurve 3}
\end{figure}

\section{Conclusions}
\label{Sect 5}
In summary, the present 15 sessions of intranight optical monitoring
of 3 $\gamma$-ray detected radio-loud NLSy1 galaxies have
raised a couple of somewhat unexpected but significant issues. One of
them relates to the importance of correcting the optical light curves
for the substantial, if not dominant thermal optical
emission. This emission is contributed by not just the host galaxy, but more
particularly by the Seyfert nucleus which is known to accrete at a
high Eddington rate in this class of AGN. Even a conservative
correction for these thermal contaminations, as estimated from the
optical imaging and SED analysis, has revealed that for two
well-observed optical outbursts of the NLSy1 1H 0323$+$342, the flux
doubling time of the optical synchrotron emission is $\leq$ 1 hour. A
more realistic correction could well bring the time scale significantly further
down, making it similar to the flux doubling times of the
ultra-rapid VHE ($>$ 100 GeV) flaring events which have been reported
for a few blazars, e.g., PKS
2155$-$304~\citep{Aharonian2007ApJ...664L..71A} and PKS
1222$+$216~\citep{2011ApJ...730L...8A}. Secondly, from the present
observations of the NLSy1 1H 0323$+$342, it appears that a large
radio-loudness parameter may not be an essential condition for the
occurrence of strong INOV in radio and $\gamma$-ray NLSy1 galaxies. Thirdly, we caution that
estimating the two well-known parameters which are commonly employed in the context of
INOV, namely optical polarization and radio loudness parameter can be challenging in the case of NLSy1s, due to the expected substantial, if not large thermal optical
contamination, as well as the marked faintness of their diffuse (i.e.,
un-boosted) radio components. The interesting hint emerging from the present study of an
admittedly small sample of three $\gamma$-ray detected, radio-loud
NLSy1 galaxies is  that {\it radio}
properties like polarization and, perhaps more evidently, the jet's
superluminal motion, are likely to serve as more potent diagnostic of
INOV in the case of narrow-line Seyfert1 galaxies.

\section*{Acknowledgments}
 We thank an anonymous referee for the helpful suggestions.  \par

\bibliography{references}

\label{lastpage}
\end{document}